\documentclass[preprint,aps]{revtex4}

\usepackage{graphicx}

\begin{document}

\title{Electronic Evidence of an Insulator-Superconductor Transition in Single-Layer FeSe/SrTiO$_3$ Films}
\author{Junfeng He$^{1,\sharp}$, Xu Liu$^{1,\sharp}$, Wenhao Zhang$^{3,4,\sharp}$, Lin Zhao$^{1,\sharp}$, Defa Liu$^{1}$, Shaolong He$^{1}$, Daixiang Mou$^{1}$,  Fansen Li$^{4}$, Chenjia Tang$^{3,4}$, Zhi Li$^{4}$, Lili Wang$^{4}$,  Yingying Peng$^{1}$, Yan Liu$^{1}$, Chaoyu Chen$^{1}$, Li Yu$^{1}$, Guodong  Liu$^{1}$,  Xiaoli Dong$^{1}$, Jun Zhang$^{1}$, Chuangtian Chen$^{5}$, Zuyan Xu$^{5}$,  Xi Chen$^{3}$, Xucun Ma$^{4,*}$,  Qikun Xue$^{3,*}$, and X. J. Zhou$^{1,2,*}$
}

\affiliation{
\\$^{1}$National Lab for Superconductivity, Beijing National Laboratory for Condensed Matter Physics, Institute of Physics,
Chinese Academy of Sciences, Beijing 100190, China
\\$^{2}$Collaborative Innovation Center of Quantum Matter, Beijing, China
\\$^{3}$State Key Lab of Low-Dimensional Quantum Physics, Department of Physics, Tsinghua University, Beijing
100084, China
\\$^{4}$Beijing National Laboratory for Condensed Matter Physics, Institute of Physics,
Chinese Academy of Sciences, Beijing 100190, China
\\$^{5}$Technical Institute of Physics and Chemistry, Chinese Academy of Sciences, Beijing 100190, China
}
\date{January 28, 2014}

\begin{abstract}
In high temperature cuprate superconductors,  it is now generally agreed that the parent compound is a Mott insulator and superconductivity is realized by doping the antiferromagnetic Mott insulator. In the iron-based superconductors, however, the parent compound is mostly antiferromagnetic metal, raising a debate on whether an appropriate starting point should go with an itinerant picture or a localized picture.  It has been proposed theoretically that the parent compound of the iron-based superconductors may be on the verge of a Mott insulator, but so far no clear experimental evidence of doping-induced Mott transition has been available.  Here we report an electronic evidence of an insulator-superconductor transition observed in the single-layer FeSe films grown on the SrTiO$_3$ substrate. By taking angle-resolved photoemission measurements on the electronic structure and energy gap, we have identified a clear evolution of an insulator to a superconductor with the increasing doping. This observation represents the first example of an insulator-superconductor transition via doping observed in the iron-based superconductors. It indicates that the parent compound of the iron-based superconductors is in proximity of a Mott insulator and strong electron correlation should be considered in describing the iron-based superconductors.
\end{abstract}


\maketitle

\newpage

The iron-based superconductors\cite{JohnstonReview,FeSCReview,StewartReview} represent the second class of high temperature superconductors after the discovery of the first class of high temperature cuprate superconductors. It is now generally agreed that the parent compound of the cuprate superconductors is a Mott insulator because of the strong electron correlation, and superconductivity is realized by doping the antiferromagnetic Mott insulator\cite{DopingMott}. In the iron-based superconductors, however, the parent compounds mostly exhibit a poor metallic behavior with an antiferromagnetic order, thus raising an issue on the extent of electron correlation and whether the picture of doping a Mott insulator is relevant to the iron-based superconductors\cite{Mott KFeSe,Mott transition}.  A related important issue under debate is whether an appropriate starting point in describing the iron-based superconductors should go with an itinerant picture or a localized picture\cite{nesting,Mott transition,Dai jianhui,wengpaper}. The observation of metallic parent compounds in the FeAs-based materials seems to favor the itinerant picture, while the insulating behavior of the A$_2$Fe$_4$Se$_5$ phase in the  A$_x$Fe$_{2-y}$Se$_2$(A=K,Cs,Ru and Tl,etc.) system suggests a possible localized nature\cite{chenxiaolong,M.H.Fang,MYi}.  Theoretical calculations indicate that the iron-based superconductors may be in proximity to a Mott insulator\cite{Mott transition,Mott KFeSe}. However, so far no clear experimental evidence of doping-induced insulator-superconductor transition has been reported in the iron-based superconductors.

The latest discovery of possible high temperature superconductivity in the single-layer FeSe films grown on a SrTiO$_3$ substrate  has attracted much attention both experimentally\cite{xue,liu,dopingpaper,DLFeng,JWang,CWChu,DLFengN,ZXShen}and theoretically\cite{theoryLiu,theoryXiangYY,theoryBazhirov,theoryZheng,theoryCao}. This system has a simple crystal structure that consists of a single-layer Se-Fe-Se unit which is an essential building block of the iron-based superconductors\cite{xue}. The superconducting single-layer FeSe/SrTiO$_3$  also possesses a simple electronic structure that exhibits only electron pockets near the Brillouin zone corner but without any Fermi crossing near the zone center\cite{liu,dopingpaper,DLFeng}. In particular, with an annealing in vacuum to change the doping level, the system undergoes a transformation from an initial N phase to the final S phase\cite{dopingpaper}. It was found that the as-grown single-layer FeSe/SrTiO$_3$ film with dominant N phase is insulating, only after a sufficient annealing can it convert into the S phase and become superconducting\cite{wenhao,dopingpaper}. This provides a good opportunity to investigate the possible insulator-superconductor transition with doping in this system.

In this paper, we report the first observation of an insulator-superconductor transition in the iron-based superconductors by performing systematic angle-resolved photoemission measurements on the single-layer FeSe/SrTiO$_3$ at various doping levels.  At a very low doping, the spectral weight near the Fermi level is suppressed, accompanied with the opening of an insulating energy gap. When the doping increases, the insulating gap decreases in size with the formation and sharpening of the coherent peak at the Fermi level. Eventually when the doping level increases to a critical value, the insulating gap closes and superconductivity starts to emerge.  The overall doping evolution in the single-layer FeSe/SrTiO$_3$ is  quite similar to the insulator-superconductor transition observed in the cuprate superconductors\cite{PYY,Yoshida}. Our observations have established the first case in the iron-based superconductors that an insulating parent compound can be doped into a superconductor. These results indicate that the parent compound of the iron-based superconductors is in proximity of a Mott insulator and strong electron correlation should be considered in describing the iron-based superconductors.

The as-grown single-layer FeSe/SrTiO$_3$  films were prepared by the molecular beam epitaxy (MBE) method\cite{xue}. Angle-resolved photoemission (ARPES) measurements were performed on our lab system equipped with a Scienta R4000 analyzer and a Helium lamp with a photon energy of 21.218 eV as the light source\cite{LiuIOP}. The as-grown samples were consecutively annealed in vacuum at different temperatures and for different times as described before\cite{dopingpaper}.  During the annealing process, it was found that two different phases can be identified in the single-layer FeSe/SrTiO$_3$ film by their distinct electronic structures:  The initial N phase in the as-grown sample  possesses an electronic structure which bears much resemblance to that  observed in the parent compound of BaFe$_2$As$_2$ in its magnetic state\cite{dopingpaper,liuPRB}, and the final S phase in the sufficiently annealed sample shows only electron pockets near the ($\pi$,$\pi$) zone corners\cite{dopingpaper}. The N phase decreases with the vacuum annealing accompanied by an increase in the S phase; they coexist in the intermittent annealing process\cite{dopingpaper}. In this work, we will concentrate on the S phase and report an observation of an insulator-superconductor transition with doping in this S phase. Because the S phase shows only electron pockets near the M($\pi$,$\pi$) points, the doping level of the S phase can be determined by the measured area of the electron Fermi pockets (see Supplementary  and Fig. S1). Different annealing sequences lead to different doping levels of the S phase. For convenience, we will label the samples annealed at different sequences with the doping level x hereafter.

A key question we will address in this work is, as soon as the S phase emerges with annealing, whether it is  metallic, insulating or superconducting.  For this purpose, we take advantage of the photoemission matrix element effect to selectively probe the electronic structure of the S phase even when it coexists with the N phase in the intermittent annealing process.  We find that, using a proper measurement geometry, the electronic structure along a momentum cut  near M2(-$\pi$,$\pi$) is dominated by the signal of the S phase, while the signal along a momentum cut near M3(-$\pi$,-$\pi$) is dominated by the N phase (see Fig. S2 in Supplementary). This makes it possible for us to focus on the signal of the S phase, as presented below.

Figure 1 shows the doping evolution of the band structure for the S phase in the single-layer FeSe/SrTiO$_3$ films (Fig. 1a-g). The measurement was performed along the momentum cut shown in Fig. 1o near the M2 point at a temperature of $\sim$20 K. For comparison, the band structure of the La-doped Bi$_2$Sr$_2$CuO$_{6+\delta}$ (La-Bi2201) at various doping levels measured along the (0,0)-($\pi$,$\pi$) nodal cut (Fig. 1p) is also presented in Fig. 1h-n. For the La-Bi2201 system, it has been shown that in the heavily underdoped region, there is an insulator-superconductor transition that occurs near the doping level of $\sim$0.10: below this doping level, there is a gap near the nodal region and the entire Fermi surface is gapped\cite{PYY}.  As seen from Fig. 1,  at a very low doping, the spectral weight of the electron-like bands in the S phase of the single-layer FeSe/SrTiO$_3$ film is rather weak (Fig. 1a). It gets stronger with the increasing doping.  It is clear that the band structure evolution with doping in the S phase is quite similar to that observed in La-Bi2201\cite{PYY}.

Figure 2a shows the photoemission spectra (energy distribution curves, EDCs) of the S phase in the single-layer FeSe/SrTiO$_3$ measured at the Fermi momentum of the electron-like band (k$_{FR}$ in Fig. 1g) at $\sim$20 K.  At a doping level lower than 0.073, the corresponding EDC shows little spectral weight at the Fermi level. When the doping level increases, a peak first emerges near the Fermi level,  then gets stronger and becomes a well-defined sharp coherence peak at high doping like x=0.114 (Fig. 2a). To check on possible gap opening, we symmetrized the original EDCs in Fig. 2c. The symmetrization procedure can remove the effect of the Fermi distribution function and provides an intuitive way of discerning a gap opening\cite{MNormanPRB}. It is clear from Fig. 2c that, when the doping level is low, there is a gap opening manifested by a spectral dip at the Fermi level.  When the doping increases, the gap size, measured by the half-distance between the EDC peaks, gets smaller and closes at the doping level of 0.089. Further increase of the doping leads to re-opening of an energy gap, as seen in the samples with dopings of 0.098 and 0.114.  For comparison, Fig. 2b shows EDCs of La-Bi2201 at various doping levels and the corresponding symmetrized EDCs are shown in Fig. 2d. The S phase in the single-layer FeSe/SrTiO$_3$ shows similar behaviors in its spectral lineshape evolution with doping when compared with the La-Bi2201 system.

The two gaps observed in the S phase of the single-layer FeSe/SrTiO$_3$, one that opens at doping lower than 0.089 and the other that opens at doping higher than 0.089, exhibit different temperature dependence (Fig. 3).  The high-doping energy gap shows a clear temperature dependence; it closes above a critical temperature ($\sim$40 K in Fig. 3i for the 0.098 sample). Its gap size as a function of temperature follows a BCS-like form (Fig. 3j).  These observations, together with its strong coherence peak (Fig. 3i) and the particle-hole symmetry observed before\cite{liu,dopingpaper,DLFeng}, strongly indicate that this represents most likely a superconducting gap. On the other hand, the low-doping energy gap behaves differently in a couple of aspects. First, the EDC peak is relatively broad at all temperatures; and thermal broadening makes it even weaker at high temperature (Fig. 3g, 0.076 sample). Second, the gap size shows little temperature dependence till the highest temperature we have measured (75 K in Fig. 3h for the 0.076 sample). Fig. 3a-f show the temperature evolution of the band structure measured along the momentum cut near M2 (as shown in Fig. 1o) for the 0.076 sample; the Fermi distribution function has been divided out in order to see part of the band above the Fermi level. Over the temperature range we measured, the spectral weight near the Fermi level is suppressed signalling the opening of an energy gap.  We also note that the spectral weight above and below the Fermi level is roughly symmetric that is similar to the nodal gap observed in La-Bi2201\cite{PYY}.  Third, the gap size can be rather large, up to $\sim$50 meV for the sample with a doping less than 0.073, which is much larger than the largest superconducting gap observed so far in the single-layer FeSe/SrTiO$_3$ ($\sim$20 meV)\cite{xue,dopingpaper,DLFeng}. These different behaviors indicate that the low-doping energy gap is distinct from the superconducting gap; most likely it is an insulating gap. We have also examined the momentum dependence of this insulating gap. As shown in Fig. 2f, within the experimental uncertainty, the gap size changes little along the ``Fermi surface", 20$\sim$25 meV for the FeSe 0.076 sample at several Fermi momenta shown in Fig. 2e. This seems to be consistent with an isotropic insulating gap although we caution that the determination of the insulating gap size involves larger uncertainty because the EDC peaks are much weaker and broader than those in the superconducting samples.

The above results clearly indicate that,  at low doping,  the S phase of the single-layer FeSe/SrTiO$_3$ shows an insulating behavior. Its gap size decreases with increasing doping, and near the doping of 0.089, the insulating gap approaches zero. Right after this doping level, superconductivity starts to emerge. Such a doping evolution is schematically summarized in Fig. 4a, together with the superconducting region at higher doping\cite{dopingpaper}.  This leads to a critical doping near 0.089 at which an insulator-superconductor transition occurs in the S phase. For comparison, we also re-plot the phase diagram of the La-Bi2201 system where a clear insulator-superconductor transition has been recently identified near the doping level of 0.10 (Fig. 4b)\cite{PYY}. It is interesting to note that, the band structure (Fig. 1), the EDCs and gap opening (Fig. 2),  and the phase diagram (Fig. 4) show many similarities between the S phase of the single-layer FeSe/SrTiO$_3$ and the La-Bi2201 system.   In the La-Bi2201 system, the parent compound is an antiferromagnetic insulator. With increasing doping, the nodal gap decreases to zero at $\sim$0.10, together with the disappearance of the three-dimensional antiferromagnetism. In the single-layer FeSe/SrTiO$_3$, it has been shown that the N phase is most likely a magnetic phase\cite{dopingpaper}, but it remains unclear whether the S phase at low doping (or zero doping) is magnetic or not.

Since it is the first time to observe the insulator to superconductor transition in the single-layer FeSe/SrTiO$_3$  film, it is important to understand its origin and ask why it has been observed so far only in this single-layer FeSe/SrTiO$_3$ but not in other iron-based superconductors.  The first question to ask is whether the observed evolution with annealing is an intrinsic doping-induced phenomenon or simply due to sample quality change or disorder effect because of the annealing procedure.  We believe the observed effect is unlikely due to sample quality change because as we know, the sample quality improves with the first few annealing sequences and degrades after too many annealing cycles\cite{dopingpaper}. However, we still identify insulating behavior in the high quality sample like 0.087 (Fig. 1) and clear superconducting behavior in the last few annealing sequences where the sample is on the verge of collapse\cite{dopingpaper}. In addition,  simple sharpening of energy bands will not alter the absence or presence of a gap and its size. On the other hand, the doping increase is clear during the annealing process (see Supplementary Fig. S1).  Another issue is whether the insulating behavior of the S phase at low doping could be caused by the presence of the coexisting N phase which is most likely magnetic and insulating\cite{dopingpaper}. It remains to be investigated about the relationship between the S phase and the N phase, and whether there is a phase separation in a real space.  Therefore, while we cannot rule out completely the possibility, the observation of an insulating gap in a sample with a doping as high as 0.087 where the S phase becomes dominant over the N phase seems to indicate that the insulating gap in the S phase is less likely induced by the N phase.

In principle, there are various ways to dope an insulator into a metal or a superconductor.  These approaches can be primarily divided into two groups: dope a band insulator, or dope a Mott insulator\cite{TokuraRMP}. In the case of the single-layer FeSe/SrTiO$_3$, since the parent compound contains an even number of 3d electrons per Fe, a natural possibility for the insulating phase is a simple band insulator\cite{Mott transition}. Doping a noninteracting (or weak interacting) band insulator with electrons can be further simplified as adding carriers into the conduction band which results in a rigid band shift.  On the other hand, the parent insulating phase can also be a Mott insulator if the electron correlation is sufficiently strong. The robustness of the Mott transition has been expected by a slave-spin approach\cite{Mott transition} and it has been proposed that the iron-based superconductors may be on the verge of the doped Mott insulator\cite{QMSi,Dai jianhui}. In fact, the iron-deficient A$_2$Fe$_{4}$Se$_5$  phase has been predicted to be a Mott insulator, and experiments have shown that it exhibits a three-dimensional antiferromagnetic transition with a much high magnetic transition temperature and a large magnetic moment\cite{WBao}. Furthermore,  it has been shown to undergo an insulator-metal transition under high pressure\cite{LLSun}. The orbital-selective Mott transition has also  been  proposed and tested in the superconducting A$_x$Fe$_{2-y}$Se$_2$\cite{QMSi,ZXShen}.  As already reported earlier\cite{dopingpaper} and further confirmed here,  the band evolution with doping in the single-layer FeSe/SrTiO$_3$ film is not a rigid band shift. On the contrary, the doping evolution of its electronic structure shows clear similarities to that of the La-Bi2201 system which agrees with a strong electron correlation picture that a Mott insulator is doped into a superconductor. This is also consistent with the theoretical expectation that the electron correlation in the iron-chalcogenides is stronger than that in the iron-pnictides\cite{ZPYin}. It is interesting to investigate whether the highly two-dimensional nature of the single-layer FeSe/SrTiO$_3$ may further enhance the electron correlation.

In summary, by systematic study on the doping evolution of the S phase in the single-layer FeSe/SrTiO$_3$ films, we have observed the first example of an insulator-superconductor transition in the iron-based superconductors. The doping evolution shows strong similarity to the insulator-superconductor transition observed in the high temperature cuprate superconductors.   These observations indicate that the iron-based superconductors are on the verge of a doped Mott insulator, and underlies the importance of including the strong electron correlation in describing the iron-based superconductors.  It is  interesting to further investigate why the single-layer FeSe/SrTiO$_3$  film shows stronger electron correlation than other iron-based compounds, and the possible interplay between the strong electron correlation  and its high temperature superconductivity.

$^{\sharp}$These people contribute equally to the present work.

$^{*}$Corresponding authors: XJZhou@aphy.iphy.ac.cn, qkxue@mail.tsinghua.edu.cn, xcma@aphy.iphy.ac.cn

\vspace{3mm}

\noindent {\bf Acknowledgement} We thank Q. M. Si and J. P. Hu for helpful discussions.  XJZ thanks financial support from the NSFC (11190022,11334010 and 11374335) and the MOST of China (973 program No: 2011CB921703 and 2011CBA00110). QKX and XCM thank support from the MOST of China (program No. 2009CB929400 and  No. 2012CB921702).

\newpage

\begin{figure*}[tbp]
\begin{center}
\includegraphics[width=1.0\columnwidth,angle=0]{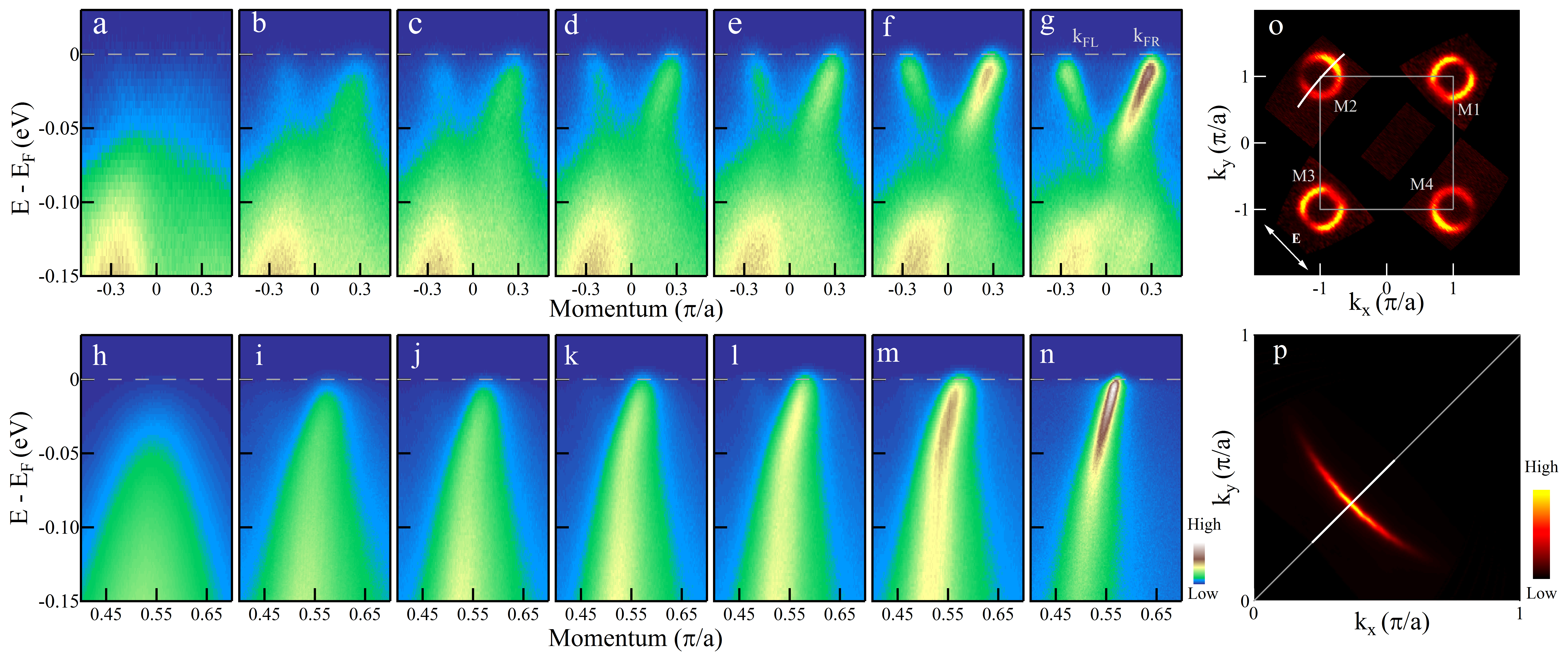}
\end{center}
\caption{Doping evolution of band structure of the S phase in the single-layer FeSe/SrTiO$_3$ film and the comparison with that in La-Bi2201. (a-g). Band structures of the S phase corresponding to doping levels of $<$0.073, 0.073, 0.076, 0.087, 0.089, 0.098, 0.114, respectively. The measurements were performed at a temperature of $\sim$20 K along a momentum cut near M2, as shown in (o) on a typical Fermi surface mapping of the 0.114 sample\cite{dopingpaper}. For comparison, the band structure evolution of La-Bi2201 is shown in (h-n) which correspond to doping levels of p=0.03, 0.04, 0.055, 0.07, 0.08, 0.10, 0.16, respectively. The measurements were performed at a temperature of $\sim$15 K along a nodal momentum cut as shown in (p) on a Fermi surface mapping of La-Bi2201 with p=0.16\cite{PYY}.
}
\end{figure*}

\begin{figure*}[tbp]
\begin{center}
\includegraphics[width=1.0\columnwidth,angle=0]{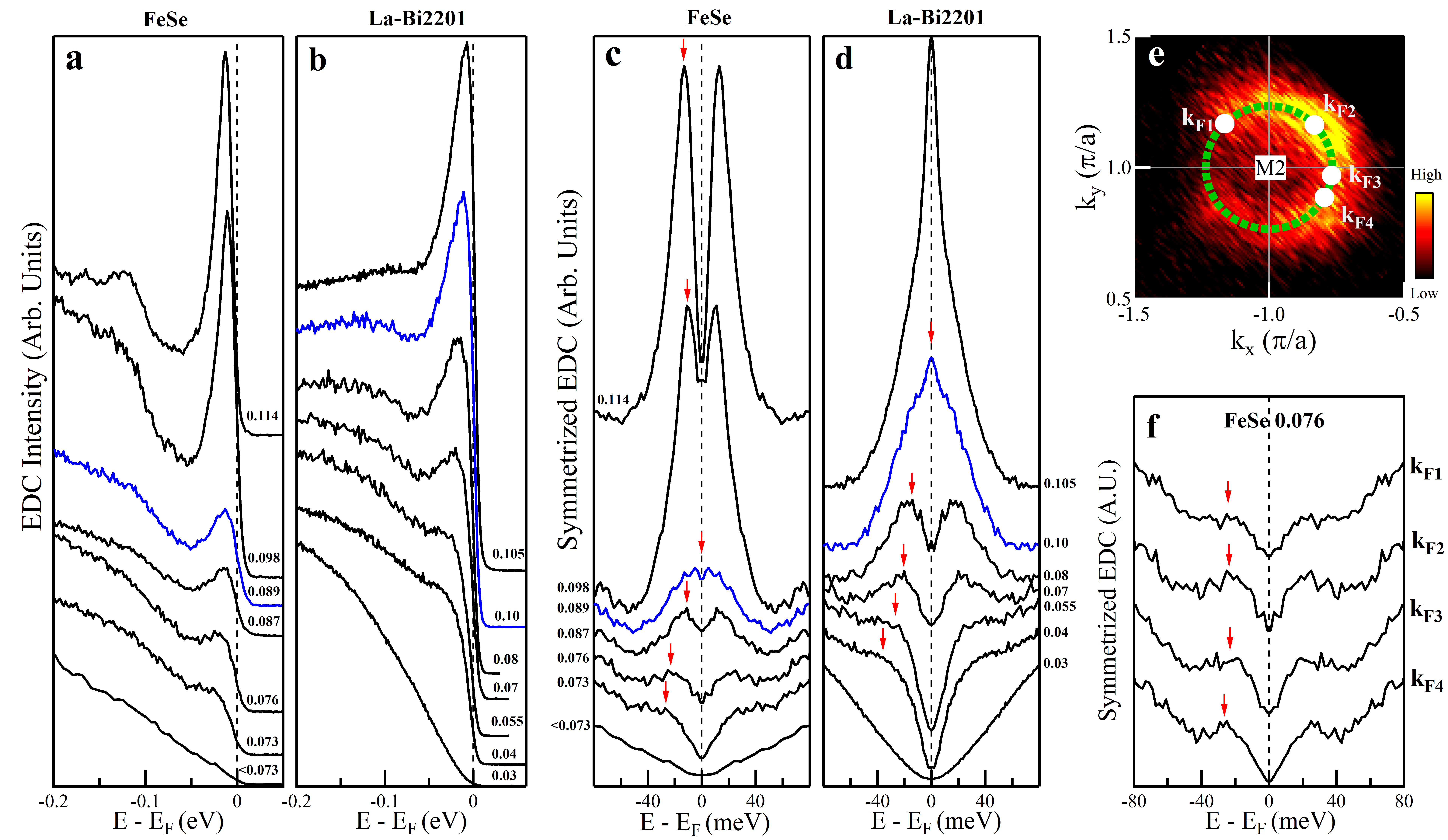}
\end{center}
\caption{Doping evolution of the photoemission spectra (EDCs) and the energy gap of the S phase in the single-layer FeSe/SrTiO$_3$ film and its comparison with that in La-Bi2201 system.  (a). Original EDCs at the Fermi momentum (k$_{FR}$) shown in Fig. 1g along the momentum cut shown in Fig. 1o for the S phase. The corresponding symmetrized EDCs are shown in (c). The EDCs are labeled by the corresponding doping levels. (b). Original EDCs at k$_F$ for La-Bi2201 along the momentum cut of the nodal direction (Fig. 1p). The corresponding symmetrized EDCs are shown in (d)\cite{PYY}. Below the doping level of $\sim$0.10, there is an opening of the nodal insulating gap.  Above $\sim$0.10, the La-Bi2201 sample becomes superconducting with a d-wave superconducting gap. No gap is observed for the 0.105 sample because the superconducting gap size is zero along the nodal direction. (e). Fermi surface mapping near M2 point for the FeSe 0.076 sample. The dashed green circle schematically shows the underlying Fermi surface.  (f). Symmetrized EDCs measured on the four Fermi momenta (marked by white solid circles in (e)) on the underlying Fermi surface in (e). Red arrows indicate the position of the energy gap.
}
\end{figure*}

\begin{figure*}[tbp]
\includegraphics[width=1.0\columnwidth,angle=0]{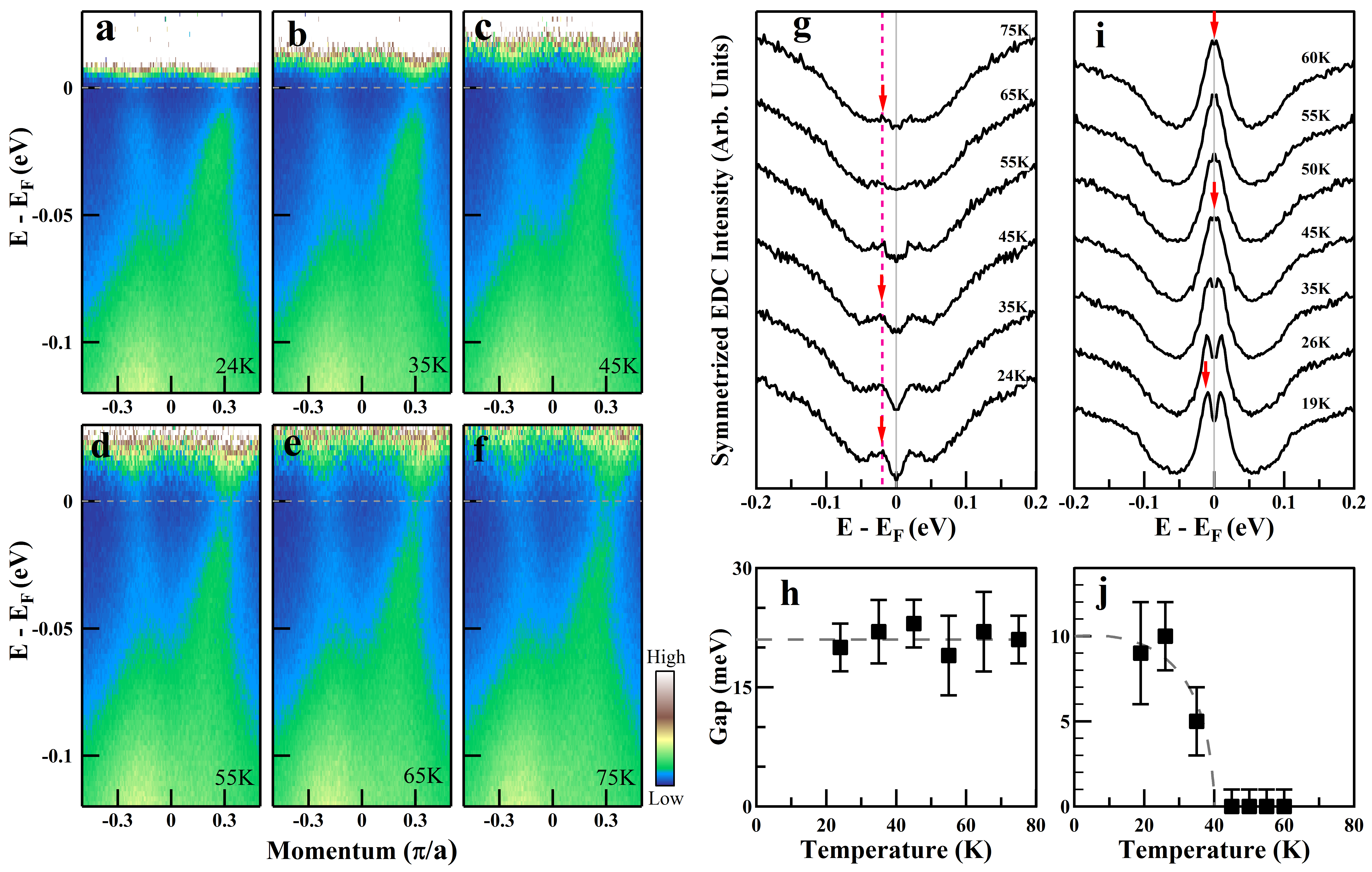}
\begin{center}
\caption{Different temperature dependence of the two energy gaps for the S phase in the single-layer FeSe/SrTiO$_3$ film. (a-f). Photoemission intensity plots of the FeSe 0.076 sample measured along the momentum cut near M2 shown in Fig. 1o at 24 K, 35 K, 45 K, 55 K, 65 K and 75 K, respectively. In order to observe part of the band above the Fermi level, the images are divided by the corresponding Fermi distribution function at different temperatures. A gap opening can be seen from the suppression of the spectral weight at the Fermi level which persists from 24 K to 75 K.  (g). Symmetrized EDCs at the Fermi momentum k$_F$  measured on the FeSe 0.076 sample at various temperatures. The red arrows indicate the position of the energy gap. The variation of the energy gap at different temperatures is shown in (h). Over the temperature range we measured, the gap size shows little change with temperature.   (i). Symmetrized EDCs at the Fermi momentum k$_F$  measured on the FeSe 0.098 sample at various temperatures.  The variation of the gap size at different temperatures is shown in (j). It decreases with increasing temperature and closes above 45 K. The gap variation with temperature follows a BCS-like form (dashed line).
}
\end{center}
\end{figure*}

\begin{figure*}[tbp]
\begin{center}
\includegraphics[width=1.0\columnwidth,angle=0]{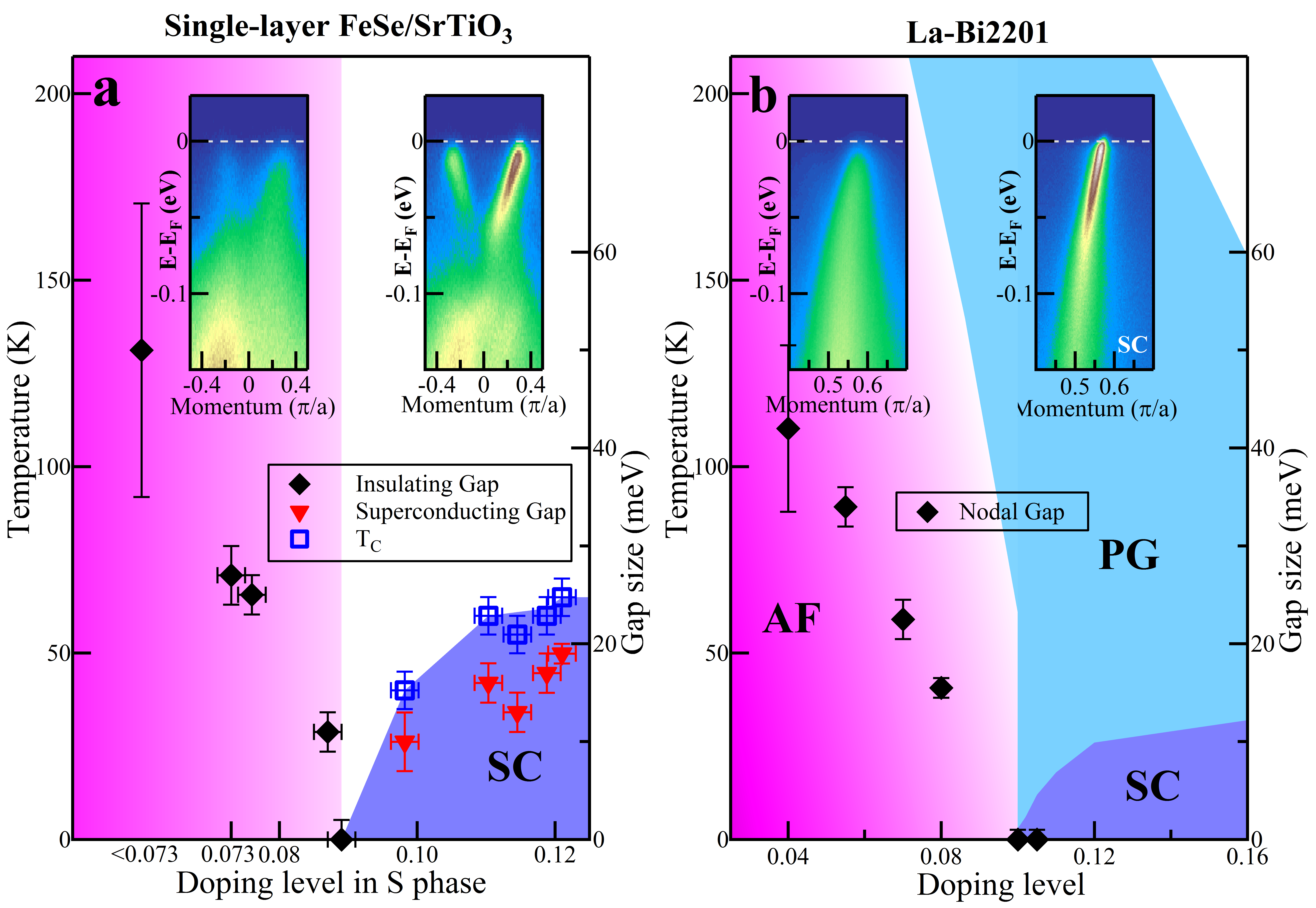}
\end{center}
\caption{Schematic phase diagrams of the S phase in the single-layer FeSe/SrTiO$_3$  film and La-Bi2201. (a). Phase diagram of the S phase in the single-layer FeSe/SrTiO$_3$  film that shows the decrease of the insulating energy gap (black solid diamond)  with increasing doping at low doping side and the increase of the superconducting gap (blue empty square) and the corresponding superconducting transtion temperature T$_c$ (red solid triangle) with increasing doping at high doping side. There is an insulator-superconductor transition near $\sim$0.09 doping level.   (b). Phase diagram of La-Bi2201 showing a clear insulator-superconductor transition near 0.10 doping level. On the lower doping side, the nodal gap (black solid diamond) decreases with increasing doping and approaches zero at $\sim$0.10\cite{PYY}.
}
\end{figure*}

\end{document}